\documentclass[aps,pra,twocolumn,groupedaddress,showpacs]{revtex4}
\usepackage{amsmath,amssymb}
\usepackage[dvips]{graphicx}

\begin{document}
\title{Enhanced Quantum Sensitivity in a Vibrating Diatomic Molecule due to
Rotational Amendment}

\author{Suranjana Ghosh,\footnote{
sghosh@iitp.ac.in} and Utpal Roy,\footnote{uroy@iitp.ac.in}}
\affiliation{Indian Institute of Technology Patna, Patna 800013,
India}

\begin{abstract}
Quantum sensitivity is an important issue in the field of quantum
metrology where sub-Planck scale structures play crucial role in
the Heisenberg limited measurement. We investigate the mesoscopic
superposition structures, particularly for well-known cat-like and
compass-like states, in the rotating Morse system where sub-Planck
scale structures originate in the dynamics of a suitably
constructed SU(2) coherent state. A detail study of the
sensitivity analysis reveals that rotational coupling in the
vibrational wave packet can be used as a probe to enhance the
sensitivity limit in a diatomic molecule. The maximum sensitivity
limit is identified with the rotational amendment, and a
quantitative measure of the angle of rotation for different
rotational levels is also given. The correspondence of the
numerical result with the angle of rotation is also delineated in
phase-space Wigner representation.

\end{abstract}

 \maketitle

\section{Introduction}

Improvement in parameter estimation has often led to scientific
breakthroughs and technological advancement. Recent advances in
experimental techniques allow us access to unprecedented levels of
control over quantum systems. Quantum metrology is the field which
exactly deals with the fundamental limits to measurement
\cite{maccone}. To reach the ultimate sensitivity limit, one can
repeat the measurement process $N$ times and take the average over
the outcomes. It reduces the error which scales as $1/\sqrt{N}$,
known as the standard quantum limit. This is the ultimate limit
one can reach using classical properties. To push this boundary,
one needs the help of quantum properties. In quantum metrology,
special states, such as entangled or squeezed states, have been
employed for estimation of these parameters to beat the standard
quantum limit \cite{maccone,caves}. In this case, the sensitivity
can be enhanced $\sqrt{N}$ times and can reach the Heisenberg
limit. On the other hand, the Planck scale executes a fundamental
role in quantum mechanics. Phase-space quasiprobability
distributions of certain quantum superposition states reveal
structures on a scale that is smaller than the Planck dimension.
The existence and importance of these small structures (called
sub-Planck structures) were first pointed out in Ref.
\cite{zurek}. These smallest interference (sub-Planck scale)
structures play a crucial role in high-precision parameter
estimation and Heisenberg limited measurement. Recently,
sub-Planck scale structures have drawn considerable attention and
have been found in different situations
\cite{pathak,toscano,dalvit,ghosh,manan,jay,sghosh1,roy,sghosh2,sghosh3,sghosh4}.
In all these studies, the sub-Planck interference phenomena appear
with a suitable combination of appropriate superposition of
coherent states (CSs) \cite{perelomov}. In our earlier studies, we
found the existence of sub-Planck structures in a molecular system
and showed their decoherence sensitivity \cite{ghosh,sghosh1}. It
involved the vibrational motion of a diatomic molecule described
by the evolution of a suitable wave packet. Governed by a
nonlinear energy spectrum, the initial wave packet breaks into
mesoscopic superpositions at a later time and gives rise to
sub-Planck structures in phase space.

In recent years, vibrational dynamics of diatomic molecules has
gained importance due to its potential application in quantum
computation. For example, ultrafast Fourier transforms can be
performed using a femtosecond laser-driven molecule \cite{r6,r5}.
High-precision molecular wave-packet interferometry has been used
to read and write the amplitude and phase information of wave
functions \cite{r2,r3}, which is a vital task for quantum
information processing and the development of quantum gates.

A key concern, however, is the effect of rotational coupling on
the vibrational motion of diatomic molecules. Our goal is to study
the sensitivity limit vis-$\acute{a}$-vis sub-Planck structures in
phase space by introducing the rotational coupling with the
vibrational motion of a diatomic molecule. Visualization of the
rovibrational dynamics needs a three-dimensional scenario
\cite{banerji,cao,lohm}, hence it is difficult to observe its
dynamics in phase space which will require a six-dimensional
configuration. To unravel this intricacy, we recall an appropriate
model, called the rotating Morse system, which can describe the
rotational and vibrational coupling nicely in one-dimensional
symmetry \cite{burkhardt,duff,gordon}. The energy eigenstates of a
rotating Morse system in phase space are elucidated in Ref.
\cite{stanek}. To the authors' knowledge, there is no study about
the wave-packet dynamics of this system in phase space. In this
study, we choose a model \cite{gordon} where the effective
potential becomes minimum around a certain equilibrium
internuclear distance, which is a function of the rotational
quantum number \textit{j}. This can satisfactorily describe the
coupling between the two degrees of freedom, i.e., the
rovibrational interplay. This coupling is also captured in
mesoscopic superpositions states, such as cat-like states and
compass-like states \cite{tara,schleich}. The most sensitive
structures in phase space, called sub-Planck scale structures, are
found to determine the sensitivity limit of a quantum state. We
have chosen the example of an iodine molecule ($I_2$), which is a
uniquely suited seed molecule for laser-induced fluorescence and
an appropriate CS wave packet is constructed to see the system
dynamics. We show the amendment in vibrational wave-packet
dynamics due to the presence of rotational coupling. Significant
advances have been made in manipulating and controlling rotational
population in rovibrational wave packets by using shaped
femtosecond pulses \cite{r7} and wave-packet interference
\cite{r1}. To make the effect of rotational coupling transparent,
we have considered a single rotational level for a rovibrational
wave packet \cite{lohm}. The sub-Planck dimension in mesoscopic
superposition structures is found to vary with the rotational
quantum number $j$. Maximum sensitivity is achieved for a
particular value of $j$ and a scheme is provided to find the exact
orientation of the corresponding system in phase space.
Additionally, the corresponding phase-space Wigner distribution is
numerically calculated and delineated in phase space to further
verify the orientation and the structures in the maximum sensitive
state.

The paper is organized as follows. We present a brief overview of
$1$D rotating Morse system and its validity. We construct the
corresponding SU$(2)$ CS wave packet to analyze the dynamics of
the CS and to explain the effect of rovibrational coupling in
configuration space. In Sec. III, we study the rotational
sensitivity in a vibrating diatomic molecule through mesoscopic
superposition structures. Specifically, we have focussed on
cat-like and compass-like states where the sensitivity issue is
explored at the sub-Planck level. The scaling law is verified and
the maximum sensitivity limit is achieved for rotational
amendment. A quantitative measure of the angle of rotation in
phase space is also depicted. Furthermore, a numerical study shows
the phase Wigner distribution, which reveals a nice correspondence
with the angle of rotation. Finally, we end up with some
conclusions in Sec. IV.

\section{Dynamics of Coherent state in Rotating Morse potential}
We start with the effective potential $V_{e\!f\!f}(r)$, known as
the rotating Morse potential,
\begin{equation}\label{effpot}
V_{e\!f\!f}(r)=D[e^{-2\beta(r-r_{0})}-2e^{-\beta(r-r_{0})}]+\frac{j(j+1)\hbar^2}{2\mu
r^2},
\end{equation}
The first part describes the well known Morse potential, an
appropriate model for vibrating diatomic molecule. $D$ is the
dissociation energy of the molecule, $r_{0}$ is the equilibrium
internuclear separation, and $\beta$ is the range parameter. The
second part stands for the centrifugal contribution of rotation. A
description of the system can be obtained with a modified
equilibrium internuclear distance $r_{j}$ and a dissociation
energy $D_{j}$ \cite{gordon}. Using a semianalytical method
\cite{burkhardt}, one can find
\begin{equation}\label{rjdj}
r_{j}=r_{0}\left[1+\frac{A}{\beta^2r_{0}^2D}\right];\;\;\;D_{j}=D\!-\!A\left(1-\frac{A}{\beta^{2}r_{0}^{2}D}\right),
\end{equation}
where $A\!=\!\frac{j(j+1)\hbar^2}{2\mu r_{0}^2}$. We define
$A_{j}\!=\!\frac{j(j+1)\hbar^2}{2\mu r_{j}^2}$ and expand the
centrifugal term of Eq.(\ref{effpot}) around $r=r_{j}$. Keeping
terms upto second order, the Schr\"{o}dinger equation is solved to
obtain the eigen functions of the rotating Morse system as
\begin{equation}\label{eigenstate}
\psi_{n,j}(y)= N_{n,j} e^{-y/2} y^{s} L_{n}^{2s} (y),
\end{equation}
where the variable $y$: $y\!\!=\!\!2\lambda_{j} e^{-\beta
(r-r_{0})}\!\!=\!\!2\lambda_{j} u e^{-\beta (r-r_{j})}$
($0\!\!<y\!\!<\!\!\infty$) and $u\!=\!e^{-\beta (r_{j}-r_{0})}$.
$n$ is the vibrational quantum number, $L_{n}^{2s} (y)$ stands for
associated Laguerre polynomial,
$s=\sqrt{(c_{0}-E_{v,j})\lambda^2_{j}/c_{2}}$, and
$\lambda_{j}=\sqrt{\frac{2\mu c_{2}}{\beta^2\hbar^2}}$. Here the
constants are expressed as
$c_{0}=3A_{j}b^{2}_{j}-3A_{j}b_{j}+A_{j}$,
$c_{1}=(3A_{j}b^{2}_{j}-2A_{j}b_{j}+u D)/u$,
$c_{2}=(3A_{j}b^{2}_{j}-A_{j}b_{j}+u^2D)/u^2$, and
$b_{j}\!=\!(\beta r_{j})^{-1}$ are dependent on quantum number
$j$. Defining $\bar{\lambda}_{j}=\frac{c_{1}}{c_{2}}\lambda_{j}$,
one obtains the constraint condition,
$2s+2n=2\bar{\lambda}_{j}-1$. $N_{n,j}$ is the normalization
constant:
$N_{n,j}\!=\!\left[\frac{\beta(2\bar{\lambda}_{j}-2n-1)\Gamma{(n+1)}}{\Gamma{(2\bar{\lambda}_{j}-n)}}\right]^{1/2}$.

The rovibrational energy eigen values $E_{n,j}$ turn out as
\begin{equation}
E_{n,j}=2\frac{c_{1}}{\lambda_{j}}(n+1/2)-\frac{c_{2}}{\lambda_{j}^{2}}(n+1/2)^2+c_{0}-\frac{c_{1}^2}{c_{2}}.
\end{equation}
It is worth pointing out that in the absence of rotation,
$c_{0}\!=\!0$, $c_{1}\!=\!c_{2}\!=\!D$, the system describes a
vibrating diatomic molecule, \emph{i.e.}, the well known Morse
potential.

\begin{figure*}[htpb]\centering
\includegraphics[width=4.8in]{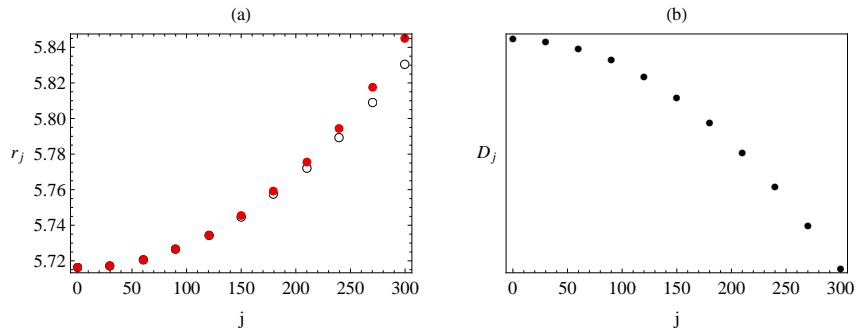}
\caption{(Color online) (a) The variation of $r_{j}$ with the
rotational quantum number $j$. It implies that the numerical
values, obtained by solving the transcendental
Eq.~(\ref{transcendental}) (red filled circles) match nicely with
the approximate values from Eq.~(\ref{rjdj}) (black circles). It
starts to differ for higher values of $j$ ($>160$). The change of
the dissociation energy of the effective potential with $j$. We
have chosen ${I_2}$ molecule where $\beta=0.9605\;a.u^{-1}$,
reduced mass $\mu=11.56\times10^{4}\;a.u.$, $r_0=5.716\;a.u.$, and
$D=0.0198\;a.u.$} \label{rjdj2D}
\end{figure*}

Alternatively, one can compute $r_{j}$ numerically by solving the
transcendental equation
\begin{equation}\label{transcendental}
\frac{dV_{e\!f\!f}(r)}{dr}\mid_{r=r_j}=0.
\end{equation}

These two sets of $r_{j}$'s are plotted in Fig.~\ref{rjdj2D}(a),
which show very good agrement for $j<160$ for $I_{2}$ molecule. It
shows that the rotational motion increases the equilibrium
distance [Fig.~\ref{rjdj2D}(a)] and decreases the dissociation
energy [Fig.~\ref{rjdj2D}(b)]. Physically, in the presence of the
rotational centrifugal force, the two constituent atoms of a
diatomic molecule tend to settle at a larger distance and are more
prone to dissociate, reducing the amount of energy required to
make them independent.

\begin{figure}[htpb]
\begin{center}
\includegraphics[width=2.6in]{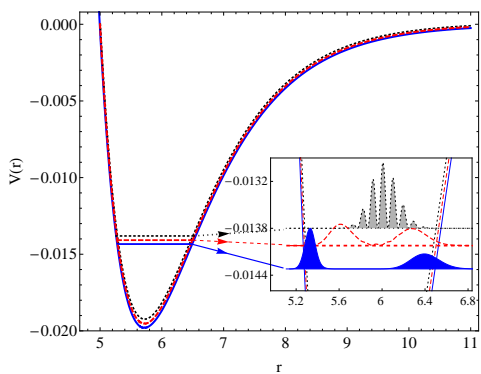}
\caption{(Color online) Effective potentials for $j=0$ (solid
line), $j=45$ (dashed line) and $j=65$ (dotted line) are depicted.
Inset: Zoom of the corresponding $10$th energy levels, where the
wave packets at $t=0.25\;T_{rev}$ are shown as dark filled
($j=0$), dashed ($j=45$) and light filled ($j=65$) plots,
respectively. The potential and the internuclear distance are in
atomic units. The corresponding parameter values are
$\beta=0.9605\;a.u.$, reduced mass $\mu=11.56\times10^4\;a.u.$,
$r_{0}=5.716\;a.u.$, $D=0.0198\;a.u.$ and $\alpha=1.6$.}
\label{2D}
\end{center}
\end{figure}

We construct a rovibrational wave packet of $I_{2}$ molecule,
which is a CS, dependent on particular rotational quantum number.
Many theoretical and experimental investigations have been carried
out on this molecule, in particular, Zewail and co-workers
investigated rovibrational wave packet dynamics in the
well-characterized electronic $B0^{+}_{u}$ state \cite{zewail1}.
Lohm\"{u}ller \emph{et al.} \cite{lohm} discussed the pump-probe
experiment of $I_{2}$ at room temperature and the detection of
fractional revivals using a full-dimensional quantum wave packet.
Here, we consider an initial rovibrational wave packet which is
centered around the $10$th vibrational energy level with $j=45$.
Under the laser polarizations magic angle conditions \cite{lohm},
it takes into account the vibrational as well as the rotational
motions. Once the bound states of the potential are included, the
dynamical symmetry group becomes $SU(2)$. For Morse system, the
corresponding SU$(2)$ generators are given \cite{dong}. In this
case, we find
\begin{eqnarray}
\hat{J_{+}}&=&\left[\frac{d}{dy}(2s-1)+\frac{1}{y}s(2s-1)-\bar{\lambda_{j}}\right]\sqrt{\frac{s-1}{s}}\nonumber\\
\hat{J_{-}}&=&-\left[\frac{d}{dy}(2s+1)-\frac{1}{y}s(2s+1)+\bar{\lambda_{j}}\right]\sqrt{\frac{s+1}{s}}\nonumber\\
\hat{J_{0}}&=&\left[y\frac{d^2}{dy^2}+\frac{d}{dy}-\frac{s^2}{y}-\frac{y}{4}+n+1/2\right].
\end{eqnarray}
$\hat{J_{0}}$ is the projection operator of the angular momentum
$m$: $m=n-\bar{\lambda_{j}}+1/2$.

We obtain the SU$(2)$ CS by operating the displacement operator
$\exp(\alpha \hat{J}_{+}-\alpha^*\hat{J}_{-})$ on the highest
bound state $n'$, defined by $\hat{J}_{+}\psi_{n',j}(y)=0$, where
$\alpha$ is the CS parameter. Temporal evolution of the CS wave
packet is given, in the eigen function basis, by
\begin{equation}\label{coherent state}
\Phi(y,t)=\sum_{n=0}^{n'}d^{j}_{n} \;\psi_{n,j}(y) e^{-iE_{n,j}t},
\end{equation}
where the weighting coefficients are evaluated as
\begin{equation}
d^{j}_{n}=\frac{(-\alpha)^{n'-n}}{(n'-n)!}\left[\frac{n'!
\Gamma(2\bar{\lambda}_{j}-n)}
{n!\Gamma(2\bar{\lambda}_{j}-n')}\right]^\frac{1}{2}.
\end{equation}
The presence of a nonlinear term in the energy expression leads to
interesting phenomena, called fractional revivals, which occur at
some specific instances between two full revivals
\cite{averbukh,robinett}. The short-time evolution displays a
classical periodicity. The classical and revival time periods are,
respectively, given by
\begin{equation}
T_{cl}=\frac{2\pi\lambda_j}{2c_{1}\!-\!c_{2}/ \lambda_j},\mbox{ and }T_{rev}=2\pi\lambda_j^{2}/c_{2}.
\end{equation}
At fractional revival times $(\bar{p}/\bar{q})T_{rev}$ (where
$\bar{p}$ and $\bar{q}$ are mutually prime integers), the wave
packet breaks into a number of subsidiary wave packets. For even
(odd) values of $\bar{q}$, the wave packet breaks into $\bar{q}/2$
($\bar{q}$) parts. In the inset of Fig.~\ref{2D}, the $10$th
vibrational energy levels for different rotational numbers are
zoomed and the rovibrational coupling effect is shown at
$t=0.25\;T_{rev}$, when CS is split into two parts. For $j=0$, the
two parts are situated at $5.3$ and $6.48$ a.u. (dark filled
plot). For $j=45$, they come close to each other, situated at
$5.62$ and $6.36$  a.u. respectively (dashed line). For a greater
value of $j$ ($j=65$), the position-space probability structure
looks completely different and shows oscillatory structure (light
filled plot). The interpretation lies in the fact that the two
split CSs oscillate inside the potential well in a back-and-forth
motion. In the first quarter of the oscillation, they approach
each other, while in the next quarter, they recede. At halfway
point of the oscillation, they are reflected from the potential
well with a phase change of $\pi$ and again become
counterpropagating. For $j=65$, they overlap each other in the
course of their oscillation and produce the oscillatory ripples,
clearly visible in the inset of Fig.~\ref{2D}. A single
interference ripple has dimension $\sim 0.1$ a.u. or $5.3$
picometers. Although, the experimental observation of small
quantum interference structures is very challenging, similar
interference ripples in picometer scale were recently visualized
experimentally for the $I_2$ molecule \cite{r4,r55}.

\begin{figure}[htpb]
\centering
\includegraphics[width=2.6in]{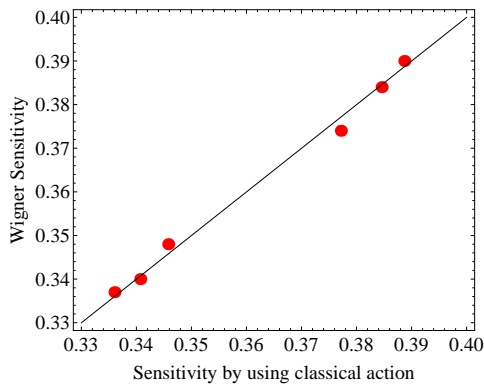}
\caption{Verification of scaling law between the sensitivity
measured from Wigner plot and numerically calculated by evaluating
the classical action. Here, the proportionality factor is $3.78$
and slope is $0.99 (\sim 1.0)$.}\label{scaling}
\end{figure}

\begin{figure*}
\centering
\includegraphics[width=6.8in]{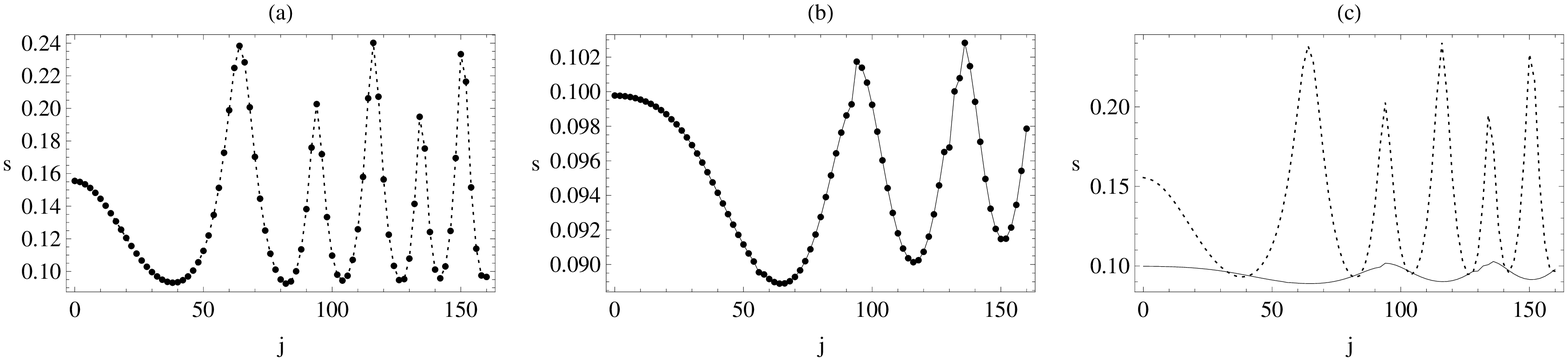}
\caption{Variation of the sub-Planck dimension with $j$: (a)
sub-Planck variation in cat-like state and (b) sub-Planck
variation in compass-like state. (c) Comparison between these two
cases.}\label{subj}
\end{figure*}
\section{Rotational sensitivity}
Until now, we have explored the wave-packet dynamics in position
space only. For a better description, we present a phase-space
picture of the dynamics. Here, we make use of the Wigner function
\cite{schleich1}, which is defined as
\begin{eqnarray}\label{wigner}
W(r,p,t)&=&\frac{1}{\pi\hbar}\int_{-\infty}^{+\infty}
\Phi'^{*}(r-r',t)\nonumber\\&& \times\Phi'(r+r',t)
e^{-2ipr'/\hbar}dr'\;.
\end{eqnarray}

\begin{figure}[htpb]
\begin{center}
\includegraphics[width=2.7in]{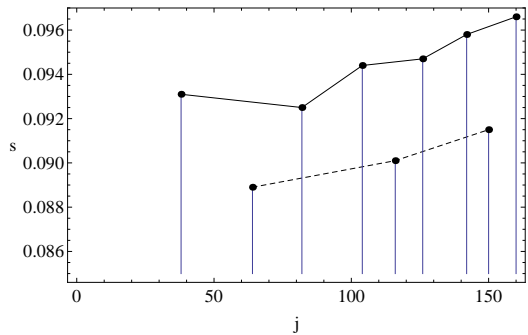}
\caption{(Color online) Points show the variation of all minima or
most sensitive sub-Planck structures in cat-like and compass-like
states. The solid line is drawn for cat-like state minima where
alternate points depict maximum sensitivity. Among them, second
minima corresponding to $j=82$ gives the most sensitive sub-Planck
region in cat-like state. The dotted line shows the compass-like
state minima variation where it brings out the most sensitivity
for $j=64$, occurs in the first minima.}\label{minall}
\end{center}
\end{figure}
Here, $\Phi'(r,t)$ is the coherent state as a function of $r$.
This Wgner function representation can reveal interesting
mesoscopic superposition structures of the CS at different times.
In addition to its positive regions,the Wigner function can also
possess negative regions for nonclassical states. In the course of
time evolution, one obtains the Schr\"{o}dinger cat-like state at
$1/4$th of the revival time. Four-way break up or the compass-like
state emerges at $1/8$th of the revival time. Sub-Planck scale
structures appear in the Wigner function at the interference
region of these mesoscopic superposition states. These structures
are alternate tiles of maxima and minima. For symmetric potentials
such as the harmonic oscillator, these tiles are rectangular in
shape and one can easily find the area of these structures by
multiplying two side arms, by measuring the distances between the
zeros of the Wigner function. However, for an asymmetric
potential, it is quite nontrivial. We find the zeros of the Wigner
function around a particular structure (either positive or
negative) by projective plots of the Wigner function in both the
conjugate coordinates. Then we perform a set of measurements and,
finally, take the average. The smallest sub-Planck structures are
formed due to the superposition of off-diagonal superposition
structures in a compass-like state.

In addition to the above procedure, one can follow an alternative
methodology, mentioned in one of our papers \cite{roy}. The idea
is as follows: In typical experimental situations, a small
perturbation can be applied through a weak constant force, which
will physically shift the state in phase space. This can be
mathematically incorporated by finding an appropriate displacement
operator for the coherent state, then applying the operator on the
state for a small displacement. When the state is displaced by the
length of a sub-Planck structure, the two states become
quasiorthogonal and distinguishable. Hence, it decides the minimum
amount of perturbation and force, which the system can detect. The
overlap between the initial and final states in terms of the
Wigner distribution is

\begin{equation}
|\left\langle\Phi'(r,t)|\Phi''(r,t)\right\rangle|^2=\frac{2}{\pi\hbar}\int_{-\infty}^{\infty}\int_{-\infty}^{\infty}W'(r,p,t)W''(r,p,t)drdp
\end{equation}
where prime denotes the state before perturbation and double
primes stand for the same after the application of an external
perturbation. One can try to find the displaced state and its
Wigner function after carrying out a lengthy calculation. The
overlap function is oscillatory and the period of each oscillation
will give the length of the structure in some particular direction
in phase space. This direction is exploited by utilizing the
complex form of the coherent-state parameters. One can, in
principle, find the length in a number of phase-space directions
to have a better idea of the shape of the structure. The method is
cumbersome and not worthy to apply in the context of the present
application because we need many of such kind of measurements. In
the following, we will investigate how rotational coupling affects
the quantum sensitivity. Specifically, we explore the sensitivity
limit due to rotational amendment in the vibrating molecule.

\subsection{Mesoscopic superposition states and their sensitivity}

In this section, we make a quantitative analysis of the
sensitivity of sub-Planck dimension with rotational coupling.
Here, we denote the dimension of the smallest structure by `$s$',
which is propotional to $\sim\hbar^2/A$ ($\sim 1/A$ in atomic
units), where $A$ is the classical action of the state in phase
space \cite{zurek,roy}. The classical action is defined by the
product of the effective support of its state in position and
momentum spaces: $A\sim \Delta x\times \Delta p$, where $\Delta
x=\sqrt{\langle x^2 \rangle - \langle x \rangle^2}$ and $\Delta
p=\sqrt{\langle p^2 \rangle - \langle p \rangle^2}$. These
quantities should be evaluated on the basis of the coherent-state
wave packet given by Eq.(\ref{coherent state}).

\begin{figure*}[htpb]
\centering
\includegraphics[width=3.5in]{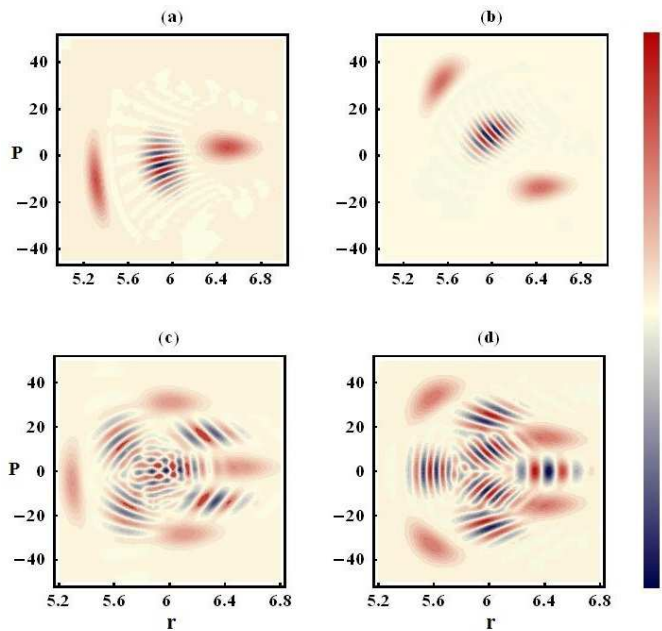}
\caption{(Color online) Time evolution of Wigner function of the
CS. The first row shows cat-like state at $t=T_{\mathrm{rev}}/4$
for (a) $j=0$, and (b) $j=82$ and the second row shows
compass-like state at time $t=T_{\mathrm{rev}}/8$ for (c) $j=0$,
and (d) $j=64$. The mesoscopic superposition of maximum
sensitivity in Rotating Morse potential}\label{wig1}
\end{figure*}

Now it brings out the question of evaluating the sensitivity or
the area of the sub-Planck scale structures. In principle, the
area can be estimated by either $(i)$ measuring the area of the
structures in the phase-space Wigner distribution or $(ii)$
measuring the classical action. The first approach needs a huge
computational time to evaluate the Wigner function integral in
each case, upon choosing a proper phase-space region. Hence, it
should be avoided, when one needs a large number of data. On the
other hand, the latter approach requires the proper scaling law
between the actual sub-Planck area and the classical action. In
our technique, we have made use of both of the approaches to
prevail over the situation. In the first step $(i)$, we plot the
Wigner function at $1/8$th revival time for only six chosen $j$
values $(j = 0, 64, 94, 116, 136, 150)$ and measure the area of
sub-Planck structure in each case as a reference value. In step
$(ii)$, we compute the inverse of classical actions for the same
parameters and, finally, compare with the reference values
obtained from Wigner plot. The scaling is depicted in
Fig.~\ref{scaling}, which produces slope $\sim 1.0$. Hence, this
is a confirmation of the scaling law: the inverse of the classical
action is directly proportional to the quantum sensitivity. Now
one would be able to perform a thorough quantitative estimate of
the sensitivity. A systematic analysis is performed to quantify
the dependence of sensitivity with different rotational angular
momentum quantum numbers and at different evolution times.

Variation of sub-Planck dimension with the rotational coupling is
shown in Fig.~\ref{subj}. As displayed in Fig.~\ref{rjdj2D}, here
we keep increasing the value of $j$ up to $160$.
Figure~\ref{subj}(a) shows the variation of the smallest
interference region with rotational amendment for a cat-like
state. Points depict the numerical values of sub-Planck dimensions
which vary with the rotational coupling parameter $j$. It is
interesting to see that the variation follows an oscillatory
behavior where all the minima represent the high sensitive
regions. With increasing value of $j$, all minima acquire
comparatively higher values. In the cat-like state, the first
minimum occurs at $j=38$. It is noteworthy to mention that the
presence of rotational coupling with particular $j$ values
corresponding to the minima shown in Fig.~\ref{subj} raise the
sensitivity limit as compared to the case for $j=0$.
Figure~\ref{subj}(b) gives the variation of sub-Planck region in a
compass-like state. In this case, numerical data shows the
oscillatory nature where all the minima capture the regions of
greater sensitivity. The first minimum corresponds to $j=64$. A
comparison is made between these two cases in Fig.~\ref{subj}(c).
It shows that the compass-like state brings out the maximum
sensitive state. A detailed study is given in Fig.~\ref{minall}.
In the cat-like state, minima or the most sensitive sub-Planck
dimensions are depicted by the points which are joined by a solid
line. It shows that $j=82$ brings out the most sensitive region in
the cat-like state. In the compass-like state, minima are joined
by a dotted line and it shows that the most sensitive sub-Planck
region arises at the first minimum corresponding to $j=64$. In the
next section, further study involves the examination of the
orientation of the system in phase space due to rotational
coupling.

\begin{table}[htbp]\centering
\caption{Angle of rotation of the wave packet corresponding to the
black dots in Fig.~\ref{minall} for both cat-like and compass-like
states.} \vskip .1in
\begin{tabular}{|rr|rr|}
  \hline
  \;\;\;\; Cat & State \;\;\; & \;\;\; Compass  &  State \;\;\;\; \\
  \hline
  \;\;\; $j$ \;\;\; \vline & \;\;$\phi$\;\;\; &  \;\;\;  $j$ \;\;\; \vline &   \;\;\; $\phi$ \;\;\; \\
  \hline
  \;\;\; $38$ \;\;\; \vline & \;\; 0.16 $\pi$ \;\; &  \;\;\; 64 \;\;\; \vline &  \;\;\;  0.22 $\pi$ \;\;\;\\
  \cline{1-2}
  \;\;\; $82$ \;\;\; \vline & \;\; 0.72 $\pi$ \;\; &  \vline  &  \\
  \cline{1-4}
  \;\;\; 104 \;\;\; \vline & \;\; 1.16 $\pi$ \;\; & \;\;\; 116 \;\;\; \vline & \;\;\; 0.72 $\pi$ \;\;\; \\
  \cline{1-2}
  \;\;\; 126 \;\;\; \vline & \;\; 1.71 $\pi$ \;\; &  \vline  &  \\
  \cline{1-4}
  \;\;\; 142 \;\;\; \vline & \;\; 2.16 $\pi$ \;\; & \;\;\; 150 \;\;\; \vline & \;\;\; 1.21 $\pi$ \;\;\; \\
  \cline{1-2}
  \;\;\; 160 \;\;\; \vline & \;\; 2.77 $\pi$ \;\; &  \vline  &  \\
  \hline
\end{tabular}
\label{table}
\end{table}

\subsection{Angle of rotation}
The rotational quantum number introduces rotation of the wave
packet in phase space and in the above section we have found the
states with maximum sensitivity for some particular values of the
rotational quantum number. Hence it is worth finding out the exact
amount of rotation $\phi$ corresponding to the states of maximum
sensitivity. Here, we provide a numerical estimation of this
rotation angle. It is well known that  $\hat{J}_0$ is the
generator of rotation and is related to the angular momentum
\cite{dong}. The corresponding rotation operator would be
$U=e^{i\hat{J}_{0}\phi}$. This operator upon operating on the
initial wave packet gives
\begin{eqnarray}\label{angle}
U \Phi(y,t)_{j=0}&=& \sum_{n=0}^{n'}d_{n}^{0}e^{i
(n-\bar{\lambda_{j}}+1/2)\phi}\psi_{n,0}e^{-iE_{n,0}t}\nonumber\\&=&\chi(y,t).
\end{eqnarray}
The resulting state is rotated by an angle, depending implicitly
on $j$.

There is a one-to-one correspondence between the above state and
the wave packet $\Phi(y,t)$, directly obtained from the time
evolution. Hence we find the angle of rotation by maximizing the
overlap $|\langle\chi(y,t)|\Phi(y,t)\rangle|^2$ for a given $j$.
Numerically estimated angles of rotation for specific important
values of $j$ are shown in Table-\ref{table}. The cat-like states
reveal maximum sensitivity for $j=82$ for which the rotation angle
is found to be $0.72\;\pi$. The angle for the most sensitive
compass-like state ($j=64$) is $0.22 \pi$.

\subsection{Phase space picture}
To obtain greater insight into what has been predicted in the
previous section, we again invoke the phase-space picture.
Figures~\ref{wig1}(a) and ~\ref{wig1}(b) display the wigner
distribution functions of the cat-like state for $j=0$ and $j=82$,
respectively. Figure~\ref{wig1}(b) clearly shows rotation of the
wave packet in phase space due to rovibrational coupling. It shows
$0.72\pi$ rotation in anticlockwise direction. Similarly,
Figs.~\ref{wig1}(c)-(d) show the wigner distribution functions of
the compass-like state for $j=0$ and $j=64$, respectively.
Following the sensitivity study, we found that the compass-like
state for $j=64$ provides the maximum precision in this rotating
Morse system. Although rotation of this particular state is
obtained as $0.22 \pi$ from Table-\ref{table}, there is another
crucial factor: the ratio of the revival and classical time
scales. This ratio is not an exact integer in most cases and the
extra fraction introduces an additional phase in the evolution.
The extra rotation is calculated to be $0.029\pi$ at
$1/8\;T_{rev}$. When added to the rotation due to $j$, the
resulting state is expected to rotate by $0.249\pi\sim\;1/4\;\pi$,
which is in conformity with the Wigner function in
Fig.~\ref{wig1}(d).

\section{Conclusion}
Proper resource accounting is crucial when investigating the
precision or sensitivity in quantum systems and formulating the
ultimate limits in quantum metrology. In this study, we have
considered the rotational coupling in the vibrating diatomic
molecule $(I_{2})$ and explored the sensitivity of mesoscopic
superposition structures. Special attention is paid to cat-like
and compass-like states where sub-Planck scale structures exist in
the quantum interference region. Our sensitivity analysis of
quantum interference structures reveals the fact that rotational
coupling enhances the sensitivity limit in a vibrating diatomic
molecule. We have also identified the rotational levels
corresponding to the maximum sensitivity limit. Our study avoids
the complication of six-dimensional phase space for rovibrational
dynamics of a diatomic molecule. The $1$-D rotating Morse
potential can well capture the rotational effect throughout the
time evolution in phase space. Moreover, we provide a quantitative
measure of the angle of rotation for different angular momentum
states. Our numerical result shows a nice correspondence between
the angle of rotation and the phase-space Wigner representation.
This study leads to an enhancement in the sensitivity limit and
hence provides improvement in the Heisenberg limit for quantum
metrology, which is not possible without rotational amendment.

\section{Acknowledgment}

The author, S. Ghosh acknowledges the support provided by DST,
Govt. of India (Fast Track project No.SR/FTP/PS-062/2010).

\end{document}